# In-operando test of tunable Heusler alloys for thermomagnetic harvesting of low-grade waste heat

F. Cugini[1,2]*, L. Gallo[1,2], G. Garulli[2], D. Olivieri[1], G. Trevisi[2], S. Fabbrici[2], F. Albertini[2], M. Solzi[1,2]

[1]Department of Mathematical, Physical and Computer Sciences, University of Parma, Parco Area delle Scienze 7/A, 43124 Parma, Italy
[2]Institute of Materials for Electronics and Magnetism - National Research Council (IMEM-CNR), Parco Area delle Scienze 37/A, 43124 Parma, Italy
*francesco.cugini@unipr.it

**Abstract**

*Thermomagnetic generation stands out as a promising technology for harvesting and converting low-grade waste heat below 100 °C. Despite the exponential growth in research on thermomagnetic materials and prototypes over the last decade, there remains, to unlock the full potential of this technology, a critical gap between fundamental research on materials and the design of advanced devices.*

*In this study, we present the in-operando assessment of thermomagnetic performance of three representative Ni,Mn-based Heusler alloys optimized for harvesting low-grade waste heat below 373 K. These materials were tested under operational conditions using a specially designed laboratory-scale prototype of a thermomagnetic motor. The mechanical power output of the motor, operated with NiMnIn, NiMnSn and NiMnCuGa alloys, was correlated with the magnetic properties of the materials, highlighting the critical role of the magnetic transition temperature and saturation magnetization in determining the efficiency of thermomagnetic energy conversion. Austenitic Heusler alloys were confirmed to be promising thermomagnetic materials due to their highly tunable Curie temperature and significant magnetization changes in the 300-360 K temperature range. Among the tested materials, the $Ni_{48}Mn_{36}In_{16}$ alloy demonstrated the highest thermomagnetic performance, surpassing the benchmark material Gd in the 320-340 K range. From an experimental perspective, the developed prototype of thermomagnetic motor serves as a flexible test-bench for evaluating and comparing the thermomagnetic performance of small amounts (less than 0.3 g) of new materials under variable conditions. Additionally, its modular design facilitates testing and optimization of its various components, thereby contributing to the advancement of thermomagnetic motor technology.*

## 1. Introduction

Technologies for energy conversion play a crucial role in modern society, impacting various aspects of our lives, including manufacturing, transportation, domestic applications, and electronic devices. However, energy conversion processes always result in significant losses, primarily in the form of waste heat released into the atmosphere. According to C. Forman et al., nearly 72% of the total energy produced by primary energy carriers in 2012 was lost as heat during conversion, with more than 45% of this loss occurring at temperatures below 100 °C [1]. The harvesting of this considerable quantity of low-grade waste heat represents a significant opportunity that could contribute to the transition to a fully sustainable and resource-efficient society by reducing the net energy consumption.

Currently, thermoelectric conversion stands as the predominant technology for the recovery of low-grade waste heat, albeit it is still characterized by low efficiencies and high costs [2,3]. Recently, thermomagnetic generation, utilizing magnetic materials, has emerged as a promising alternative, particularly for harvesting of heat from sources at temperatures below the boiling point of water (T < 373 K) [4,5]. This technology relies on thermomagnetic (TM) cycles, leveraging the temperature-dependent magnetization of a magnetic material to convert thermal energy into mechanical or electrical energy. The concept traces back to the late nineteenth century when both N. Tesla and T. A. Edison independently patented prototypes of thermomagnetic generators (TMGs) [6,7]. Nevertheless, these devices exhibited significantly low efficiency near room temperature, primarily due to the use of a magnetic material (Fe) characterized by a very high Curie temperature ($T_c$ = 1043 K). Consequently, the scientific community lost interest in thermomagnetic generation. In recent decades, advancements in the study of magnetocaloric materials for magnetic refrigeration at room temperature [8], have rekindled research interest in TM generation for low-grade heat harvesting [9,10]. Numerous proof-of-concept prototypes of TMGs have been developed, showcasing the feasibility and potential of TM generation concept [5,11]. However, to fully unlock its potential and make it economically viable for widespread use, this technological innovation still requires the identification of high-performance, sustainable, recyclable and cost-effective TM materials and their efficient integration into devices.

TMGs operate based on a thermomagnetic cycle, schematically illustrated in Figure 1. A typical thermomagnetic cycle involves two isothermal transformations, in which the magnetic field is applied (*I*) and removed (*III*), and two iso-field transformations, with the absorption of heat from a warm source (*II*) and its release to a cold source (*IV*). Both changes in temperature and field influence the magnetization of the material *M(T,H),* as depicted in Figure 1.b. The magnetic work (*$W_m$*) produced by the cycle can be computed as the net variation of Gibbs free energy and corresponds to the area enclosed by the cycle in the *M(H)* diagram (Figure 1.b) [12-14]:

$$W_m = \mu_0 \oint M \, dH \quad (1)$$

This outcome, derived from an ideal cycle, established the primary criteria for selecting optimal TM materials. An effective TM material should exhibit a substantial change in magnetization within the working temperature and field ranges of the application. The achievement of this goal relies on the selection of materials with a magnetic phase transition in the desired temperature range between two phases characterized by a significant difference in magnetization.

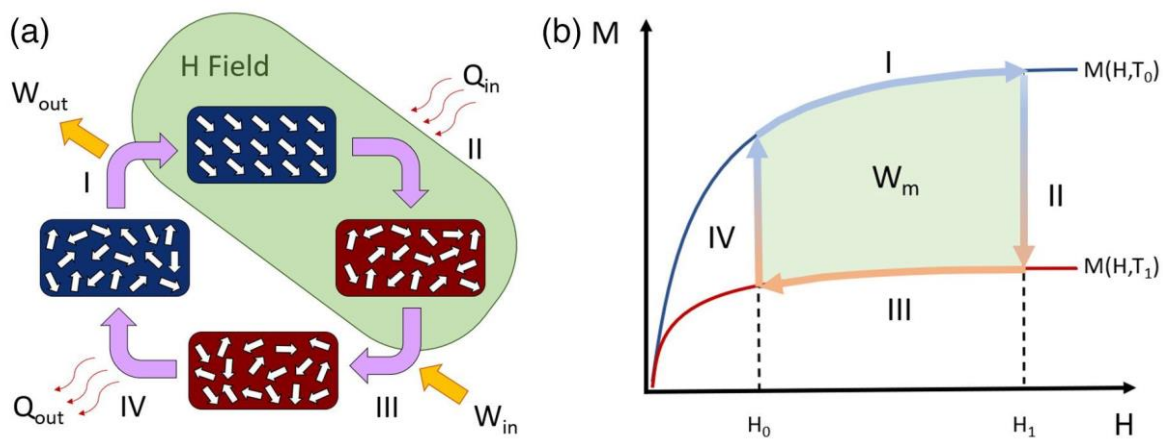

*Figure 1: (a) Conceptual scheme of the thermomagnetic cycle exploited in TMGs and (b) representation of the cycle in the M-H diagram. The cycle is enclosed between two isotherm magnetization curves at temperatures $T_0$ and $T_1$. The useful magnetic work of the cycle $W_m$ is highlighted by the green area.*

From Eq. 1, it is possible to derive the efficiency $\eta$ and the power *P* expected from an ideal thermomagnetic cycle working between thermal reservoirs at temperatures $T_0$ and $T_1$ [5,13]:

$$\eta = \frac{|W|}{Q_{in}} \cong \frac{\mu_0 \oint M \, dH}{\rho \int_{T_0}^{T_1} c_p(T) dT} \quad (2)$$

$$P = \left(\mu_0 \oint M \, dH\right) f \quad (3)$$

where $\rho$ is the density, $c_p$ the specific heat of the material and *f* the working frequency of TM cycle. Dzekan *et al.* introduced a further approximation of Eq. 3, serving as a simplified figure of merit for an initial comparison of thermomagnetic materials [10]:

$$P = \mu_0 H \Delta M f \quad (4)$$

where *ΔM* is the magnetization change between temperature $T_0$ and $T_1$ in a magnetic field $H_1$.

In addition to the necessary magnetic properties, Eq.s 2-4 highlight the significance of the thermal characteristics of active TM materials in determining the power and efficiency of a TMG. Thermal capacitance and thermal diffusivity govern the rate and the quantity of heat transferred between the material and the thermal sources, influencing the width of the TM cycle and of the maximum working frequency of the generator. [5,11]

To date, a few materials classes are considered the most promising for TMGs [10,13]. While gadolinium serves as benchmark for magnetocaloric applications near room temperature and it is utilized in the majority of TMG prototypes [15-21], its Curie temperature of 292 K is unfavourable for applications in which the cold reservoir is at room temperature (around 293 K) and the heat source operates at temperatures between 300 and 370 K. $La(Fe,Si)_{13}H_x$, MnFePSi and Heusler compounds represent broad classes of materials exhibiting both first-order and second-order magnetic transitions that can be tuned over a wide temperature range by varying the composition. These characteristics, coupled with their low criticality in terms of raw materials availability, make them highly promising materials for TM generation. [10]

In particular, Ni-Mn based Heusler alloys, with general formula $Ni_{50}Mn_{25+x}Z_{25-x}$ (where Z typically represents an element of the p-groups), are stable in a wide range of compositions, allowing for numerous combinations of elements. Consequently, this enables the fine tuning of magnetic and magnetostructural properties across a broad temperature range. [22-25]. These alloys can exhibit both a second-order Curie transition of the ferromagnetic austenitic phase and a first-order magnetostructural transition between the cubic austenite and a low-temperature and low-symmetry phase (martensite). Both transitions generate large values of magnetocaloric effect [26,27]. Furthermore, Heuslers generally have a high thermal diffusivity, are rare-earth free and predominantly composed of non-critical and low-cost elements. [8,10] The $Ni_{49}Mn_{35}In_{16}$ alloy demonstrated the maximum magnetocaloric effect at the Curie transition just above room temperature, thanks to its large saturation magnetization (approximately 125 $Am^2kg^{-1}$) [27]. Substituting In with Sn induces a small increase in the critical temperature and a significant decrease in saturation magnetization, due to changes in the electronic band structure [28,29]. Ga-based Heusler alloys can exhibit a coupling between the second-order transition and the magnetostructural transformation. [30,31]. The addition of Cu can be used to finely tune the transition near room temperature and improve the mechanical properties of the alloy [25,32-35].

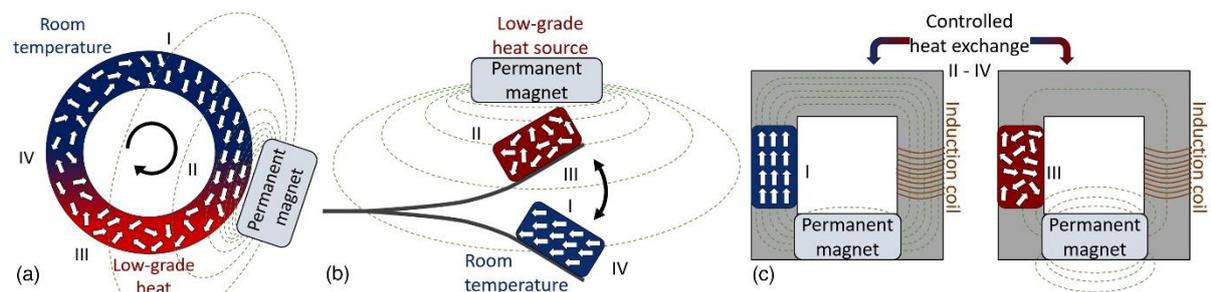

Figure 2: Sketches of (a) TM rotor, (b) TM oscillator and (c) TM static generator for the harvesting of waste heat. The TM material is colored in red or blue depending on its temperature. The arrows in the material depict the arrangement of magnetic moments inside the material. The numbers in brackets correspond to the steps of the TM cycle reported in Figure 1.

Thermomagnetic cycles find application in thermomagnetic generators. To date, three main configurations of TMGs have been successfully realized: thermomagnetic motors, oscillators and static generators. [5,11] Thermomagnetic motors (Figure 2.a), also known as "Curie wheels", utilize a rotor made of a TM material to convert thermal energy absorbed from a warm source into rotational energy. [15, 19, 36] This conversion is achieved through the rotor interaction with a magnetic field gradient. The temperature variation of each circular sector of the rotor, caused by cyclical heat absorption from the warm source and subsequent release to the environment, induces a magnetization change of the TM material. Consequently, there is a variation in the force of interaction with the magnetic field gradient, resulting in a net momentum that increases the rotational speed of the rotor. During rotation, each circular sector of the rotor undergoes the thermomagnetic cycle, sustaining the motion. The mechanical energy of the rotor can then be converted into electrical energy by a conventional alternator. Thermomagnetic oscillators (Figure 2.b) share conceptual similarities with TM motor: they convert thermal energy in vibrational mechanical energy of a flexible support. [17,19,20,37,38] Static generators, instead, convert thermal energy into electrical energy by cyclically varying the temperature of a TM material (Figure 2.c), which acts as a switch of magnetic flux inside a magnetic circuit. [39,40,41,42]

In this study, we experimentally explore the potential thermomagnetic performance of three representative Ni,Mn-based Heusler alloys, specifically optimized for applications in harvesting of low-grade waste heat below 373 K. The TM potential of these alloys is assessed through a series of magnetization measurements as a function of temperature and applied magnetic field within the working regime. In addition, their TM performance is directly evaluated in a small-scale thermomagnetic motor prototype, designed as a test-bench for in-operando assessment of thermomagnetic materials. This device allows the characterization of small masses of materials (less than 0.3 g), enabling the direct testing and comparison of materials performances on a research laboratory scale. This approach aims to bridge the gap between the study of the fundamental properties of TM materials and their integration into devices. Moreover, the study delves into the relationship between the maximum mechanical output of TM motor and the magnetic work of the exploited TM cycle, calculated from magnetization data, providing a method for evaluating and comparing the potential of magnetic materials in TMG applications. The obtained results experimentally confirmed the strong dependence of TM performance on the strength of material's variation of magnetization with temperature. This correlation underscores the importance of optimizing TM materials for each specific working temperature range. Lastly, the developed TM motor prototype is used to highlight the critical role of thermal and geometric properties of the rotor in determining the mechanical power output, paving the way to improve the performance of TM generation technologies.

## 2. Material and methods

### 2.1 *Alloys preparation*

The alloys, with nominal compositions $Ni_{48}Mn_{36}Sn_{16}$, $Ni_{48}Mn_{36}In_{16}$, and $Ni_{50}Mn_{19}Cu_6Ga_{25}$, were synthetized by arc melting under Ar atmosphere the stoichiometric amount of high-purity (99,99 %) elements. To compensate for evaporation losses, a 1% excess of Mn was added. The ingots were flipped and re-melted three times to enhance homogeneity. Subsequently, the samples were annealed at 1073 K for 72 hours and quenched into water. The composition of the samples was analysed using energy dispersive X-ray spectroscopy (EDX) in a scanning electron microscope (Bruker Esprit microanalysis on a FEG-ESEM FEI QUANTA 250). All samples compositions were found to be in agreement with the nominal ones within the 1% margin of experimental error. The prepared bulk samples were manually grinded in an agate mortar and sieved to obtain powder with a maximum size of 120 µm. The powders were annealed at 773 K for 4 h in Ar atmosphere, followed by slow cooling, to reduce the stresses introduced during grinding and to recover the magnetic properties of the bulk samples. [35,43]

### 2.2 Magnetic characterization

The temperature dependence of the low-field a.c. magnetic susceptibility of the samples was measured as a function of temperature with an applied ac magnetic field of 1 mT using a ThermoMagnetic Analyzer (TMA). This allowed us to identify the Curie temperature of the materials and ascertain the absence of other magnetic transitions. Magnetization measurements were performed as a function of temperature using an extraction magnetometer (Maglab2000 System by Oxford Instruments) and a force magnetometer (DSM8 by Manics). These measurements were used to obtain high-field magnetization values and to calculate the potential thermomagnetic performance of the materials from $M(T,H)$ data. Hysteresis loops were recorded for the three alloys at room temperature in an external magnetic field ranging between ±1T with the extraction magnetometer and are reported in Figure S2 of SM.

### 2.3 Thermomagnetic tester

The small-scale TM motor prototype designed to test TM materials under operative conditions falls under the category of "Curie wheels". Figure 3 reports a photograph and a sketch of the TM motor prototype, which primarily consist of three elements: (1) a rotor made of a TM material with a cylindrical symmetry; (2) a controlled heat source; (3) an assembly of permanent magnets generating a magnetic field gradient with a maximum field of 0.6 T (further details on the permanent magnet assembly and its stray magnetic field are reported in section 1 of SM) . The rotor is partially immersed in warm water at temperature $T_1$, simulating the heat source, and in the magnetic field gradient. Heat exchange between the rotor and the heat source, as well as with the environment at temperature $T_0$, generates an angular gradient of temperature inside the rotor $T(\theta)$. Consequently, the magnetic moment of the rotor and the magnetic force of interaction with the field gradient also vary with the angular position ($m(\theta)$, $F(\theta)$). The appropriate combination of temperature and field gradients on the rotor results in a torque, derived from the sum of tangential magnetic forces acting on each angular sector, characterized by a magnetic moment $m(\theta)$ and subjected to a field gradient in the tangential direction $\partial H(\theta)/\partial u_t$ :

$$\overrightarrow{\tau_{net}} = \vec{R} \times \mu_0 \left( \int_0^{2\pi} m(H,T,\theta) \frac{1}{R} \frac{\partial H(\theta)}{\partial u_t} d\theta \right) \overrightarrow{u_t} \qquad (5)$$

where $\vec{R}$ is the radius vector of the rotor. This torque induces the rotation of the rotor, leading to the conversion of thermal energy into mechanical energy. The temperature of the warm source can be adjusted within the 290-350K temperature range and it is stabilized within an error of 0.1 K. The water level can be adjusted to alter the relative position between the thermal and magnetic gradients. The non-immersed part of the rotor can be exposed to the room environment or to a controlled flux of thermostated air within the 280-300K temperature range. The rotor shaft is connected to a custom-designed two-phase electric generator, enabling the measurement of the thermomagnetic rotor angular speed and torque. The time-dependent angular speed is proportional to the voltage measured across the two windings of the generator, while the torque is derived from the current measured in the windings. A calibration procedure, involving the application of a known torque to the shaft, was employed to determine the constants of the electric generator and to measure the friction torque of the TM motor (further details are provided in the SM). To obtain the overall torque generated by the TM motor, all torque data collected during materials testing were corrected for the measured friction torque values. The time dependent mechanical power is calculated as the product of angular speed and torque. A controlled electronic load was employed to assess the TM motor performance by adjusting the external load. This approach enables the evaluation of the rotor performance across different rotation speed. Integrated software enables continuous measurement of the TM motor angular speed, torque and mechanical power output, along with monitoring the temperatures of the cold and warm reservoirs.

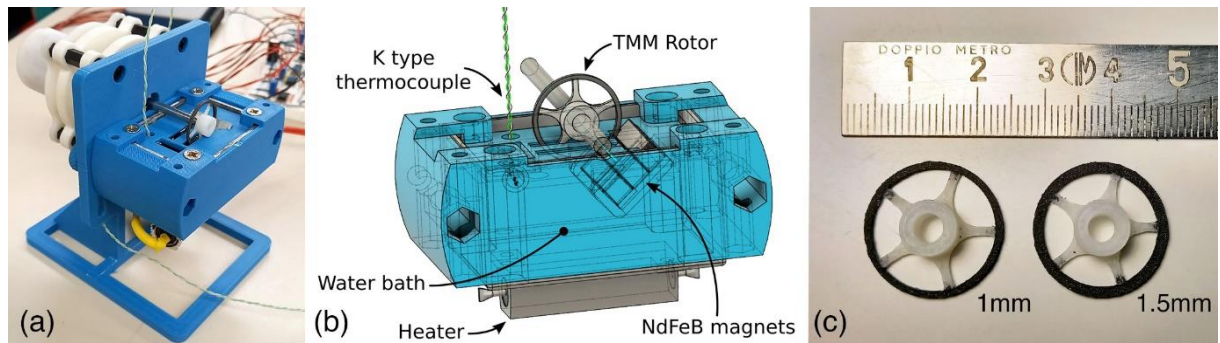

*Figure 3 (a) Photograph and (b) technical scheme of the small-scale TM motor prototype utilized for testing materials. (c) Example of rotors made of NiMnIn with a radial thickness of 1 mm and 1.5 mm.*

The active rotor (Figure 3.c) is made of an epoxy-based composite material incorporating powder of a TM material, following the procedure detailed in Ref. [444]. The ring shape is achieved by casting the composite in a mould of water-soluble bioplastic, realized with a fuse filament deposition 3D printer, that contains at the centre an insoluble plastic support for the connection to the motor shaft (Figure 3.c). The support is specially designed to minimize contact area with the active material and enhance the heat exchange of the rotor with the environment. The proposed method enables the preparation of homogeneous rotors from magnetic material powder with a particle size of less than 200 µm. The homogeneity can be assessed by measuring the time dependence of the rotational speed in the TM motor. Any significant inhomogeneity in the rotor, whether due to improper preparation procedure or deterioration during operation, is indicated by variations in the angular velocity within a single rotation cycle.

For this study, we prepared ring-shaped rotors for each of the three investigated materials (NiMnIn, NiMnSn and NiMnCuGa). The rotors have an external diameter of 20 mm, an internal diameter of 18 or 17 mm and a length of 1 mm (Figure 3.c). Utilizing a composite with 87±2 wt% magnetic powder, the net mass of active magnetic material in the rotors with an internal diameter of 18 and 17 mm is approximately 0.23±0.01 g and 0.34±0.01 g, respectively.

The evaluation of the three materials in the TM motor was performed recording mechanical power output and rotation speed while varying the resistive load and the temperature of the hot water. The load sweep was conducted in both increasing and decreasing load resistance to validate measurement repeatability. Power versus speed curves were recorded for each rotor by using as cold source a controlled flux of thermostated air at 297 K and with increasing warm water temperatures in 5 K increments from 303 K to 338 K. The rotor was submerged to a depth of 3 mm in the water. This level was selected to position the peak of the stray magnetic field at the water surface. Consequently, one peak of the magnetic field gradient is situated just above the water level, while the other peak is fully immersed in the warm water (further details on the permanent magnet assembly and its stray magnetic field are provided in the section 1 of SM). Under static conditions, this geometrical configuration of the thermomagnetic motor maximizes the torque on the rotor by optimizing the temperature difference between the rotor sections experiencing the maximum field gradient.

## 3. Results and discussion

### 3.1 Magnetic and thermomagnetic properties

Figure 4 presents low-field ac magnetic susceptibility under zero applied dc magnetic field and magnetization measurements in an applied magnetic field of 1 T and 1 mT for the three Heusler alloys under investigation. NiMnIn and NiMnSn alloys exhibit a second-order Curie transition at 317 K and 323 K, respectively. The temperature behaviour of the high-field magnetization of these alloys is similar, with the magnetization decreasing by approximately 25% upon replacing In with Sn. In contrast, the NiMnCuGa alloy undergoes two magnetic transitions within a narrow temperature range. At 301 K the sample undergoes the second-order Curie transition on heating between the ferromagnetic and paramagnetic state of austenite. At 275 K, the austenitic phase transforms on cooling to martensite, characterized by lower susceptibility and a larger high-field magnetization. The magnetization variation with temperature $dM(T)/dT$ is maximal around the critical temperatures for all samples and decreases proportionally to the value of high-field magnetization. The NiMnCuGa alloy shows a maximum $dM/dT$ at the sharp first-order transition around 275 K. However, the thermal hysteresis ($\Delta T_{hyst.}$ = 6 K) and the small temperature range in which the $dM/dT$ is sizable make this transition less useful for cyclic TM applications. A summary of the main magnetic properties crucial for TMG application of the investigated Heusler materials is provided in Table 1. The finite magnetization change calculated on a temperature span of 4 K around the Curie temperature ($\Delta M_{4K}$) and the maximum high-field magnetization derivative with temperature ($dM/dT_{max}$) follow a similar trend with the saturation magnetization of the alloys and can be equally utilized for a preliminary evaluation of potential TM performance of materials. All three alloys exhibit a very soft magnetic character at room temperature. The hysteresis loops, as reported in the paragraph 2 of the SM, show a coercive field lower than the magnetometer resolution (1 mT). This confirms the very low coercivity, which is associated with the very low magnetic anisotropy of austenitic Heusler alloys, as previously demonstrated in the literature (e.g., Ref. [29] for NiMnIn and NiMnSn alloys or Ref. [45] for the austenitic phase of NiMnGa alloys).

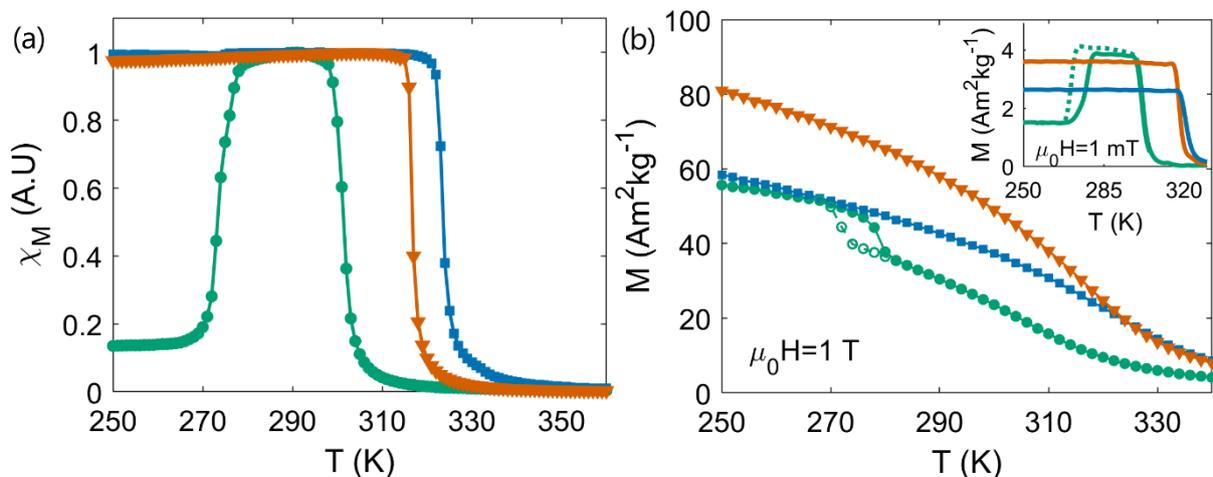

*Figure 4: (a) normalized ac magnetic susceptibility in zero applied dc magnetic field and (b) magnetization in 1T applied magnetic field as a function of temperature for the alloys: $Ni_{48}Mn_{36}Sn_{16}$ (blue squares), $Ni_{48}Mn_{36}In_{16}$ (orange triangles), and $Ni_{50}Mn_{19}Cu_6Ga_{25}$ (green circles). Inset of (b): magnetization of the three samples in an applied magnetic field of 1 mT.*

*Table 1: composition, Curie temperature derived by low-field ac susceptibility measurements, saturation magnetization, maximum temperature variation of magnetization and maximum magnetization change, in an applied magnetic field of 1 T, collected over a temperature span of 4 K around the Curie temperature. *The saturation magnetization of NiMnCuGa austenitic phase was extrapolated from M(T) data above the martensitic transition.*

| Sample | Composition | $T_C$ (K) | $M_s$ (Am$^2$Kg$^{-1}$ K$^{-1}$) | $dM/dT_{max}$ (Am$^2$Kg$^{-1}$ K$^{-1}$) | $\Delta M_{4K}$ (Am$^2$Kg$^{-1}$ K$^{-1}$) |
|---|---|---|---|---|---|
| NiMnSn | $Ni_{48}Mn_{36}Sn_{16}$ | 323 | 96.2 [28] | -0.87 | 5.3 |
| NiMnIn | $Ni_{48}Mn_{36}In_{16}$ | 317 | 135.5 [28] | -1.31 | 8.2 |
| NiMnCuGa | $Ni_{50}Mn_{19}Cu_6Ga_{25}$ | 301 | 70* [46] | -0.81 | 4.9 |

Series of isothermal *M(H)* measurements (Figure 5) were collected at various temperatures within the operational range of TM generators for the harvesting of low-grade waste heat (290-380 K). These measurements are utilized to calculate the maximum useful magnetic work per mass unit that the TM material can produce in a thermomagnetic cycle. As described in Eq. 1, the area enclosed between two isothermal *M(H)* curves at $T_0$ and $T_1$ in the $H_0$-$H_1$ field interval corresponds to the magnetic work of the cycle. Using this equation, we can assess the expected TM performance of materials based on external conditions that define the cycle boundaries. Given the very low coercive field of the studied materials and the absence of thermal hysteresis, the irreversible contribution of the magnetic transformation within thermomagnetic cycles can be neglected in the calculation. Conversely, for materials undergoing a first-order magnetic transition or characterized by large coercivity, irreversible effects must be carefully considered. Specifically, in the case of a first-order transition, attention must be paid to the development of partial transformations within minor loops if the temperature width of the thermomagnetic cycle is comparable to or lower than the thermal hysteresis [10].

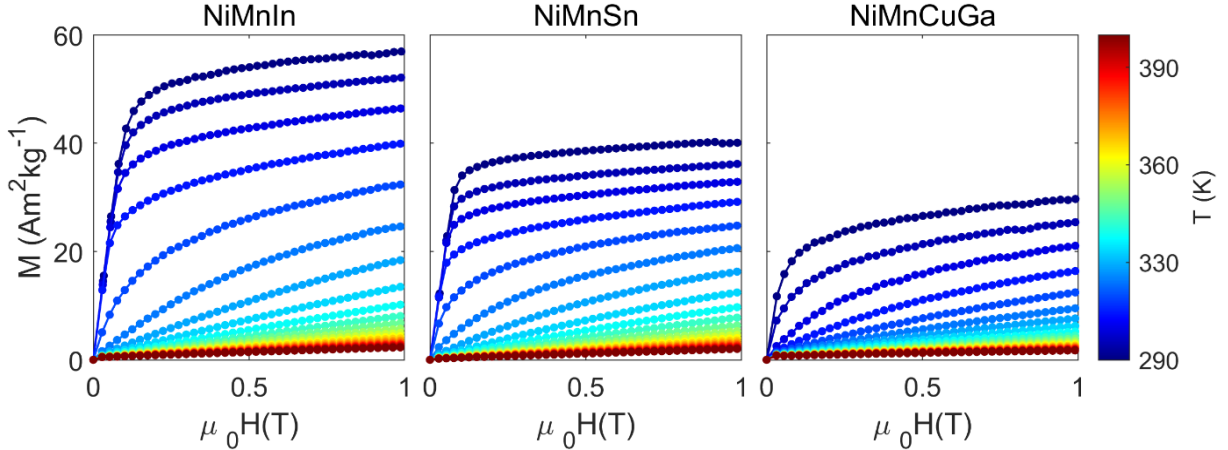

*Figure 5: magnetization as a function of external magnetic field collected at different temperatures for the NiMnIn, NiMnSn and NiMnCuGa samples.*

Figure 6.a presents a comparison of the useful magnetic work obtained for the three materials under investigation and the benchmark gadolinium. This assessment considers a maximum magnetic field of 0.5 T, a cold source at 297 K and the warm reservoir that varies from 300 K to 400 K. Near room temperature, Gd exhibits the highest potential output due to its large saturation magnetization (254 Am$^2$kg$^{-1}$ [8]), which is twice that of the best Heusler alloy (NiMnIn, $M_s$ = 135.5 Am$^2$kg$^{-1}$ [28]). Beyond 310 K, the magnetic work produced by Gd ceases to increase due to the minimal contribution of the paramagnetic phase above 292 K. At higher temperatures, above 313 K, a TM cycle operating with NiMnIn material produces more work. NiMnSn ($M_s$ = 96.2 Am$^2$kg$^{-1}$ [28]) exhibits a temperature dependence of the output magnetic work similar to that of NiMnIn, but the maximum magnetic work is reduced by about the 25% due to the decrease in saturation magnetization. NiMnCuGa shows a further reduction in maximum magnetic work, but its lower Curie temperature makes it more performant than the other Heuslers in the low-temperature range. These results highlight the significance of saturation magnetization and Curie temperature as the two key parameters in determining the optimal TM material based on the working temperature range, particularly for materials exhibiting a second-order Curie transition.

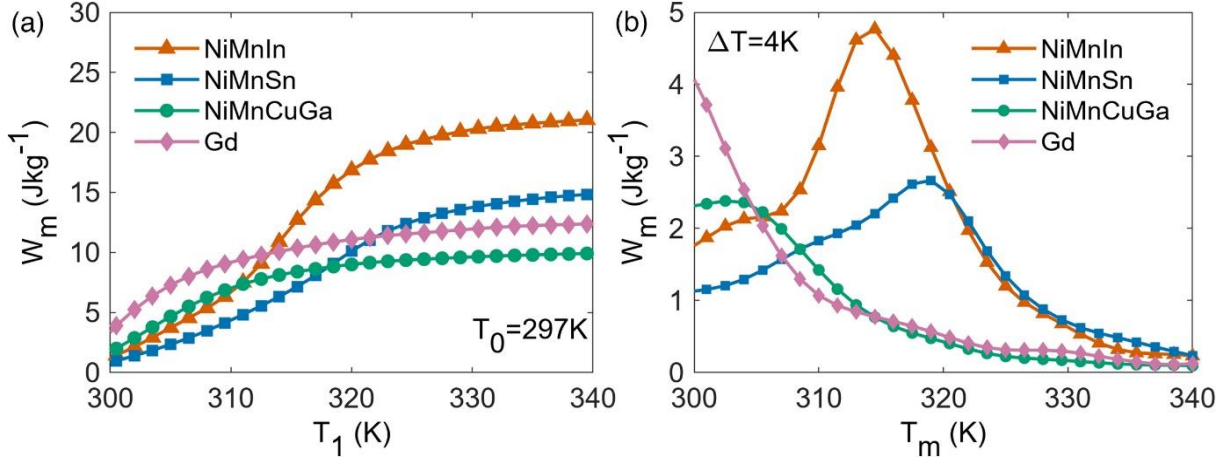

*Figure 6. (a) Useful magnetic work of an ideal TM cycle operated with NiMnIn, NiMnSn, NiMnCuGa and Gd materials as a function of the warm temperature of the cycle $T_1$. The cold temperature and the magnetic field span are considered constant at 290 K and 0.5 T, respectively. (b) Useful magnetic work as a function of the mean temperature of the cycle $T_m$, with a temperature span between the cold and warm source fixed at 4 K. The magnetic work is calculated through Eq. 1 from M(T,H) data reported in Figure 5.*

Nevertheless, using the thermal reservoirs temperatures as the integration limits for the cycle is unrealistic, given the characteristic frequencies and heat exchanges occurring in real applications. Indeed, all reported TMG prototypes indicate that the active magnetic material does not reach the temperatures of the thermal sources during the cycle, primarily due to the limited heat exchange. This limitation is mainly related to the thermal proprieties of the active material, the necessity for high working frequencies of cycling to enhance power and the design constraints of the TMG. It represents a complex challenge with no straightforward solution. Both a high working frequency and a cycle within a broad temperature span are crucial to increase the power and the efficiency of TMGs (Eq. 3). Hence, the simulation of the useful magnetic work of the cycle has been reiterated, considering an almost constant temperature change achieved by the material during the cycle and restricted by the heat exchange with the thermal sources. Figures 6.b displays the expected magnetic work calculated for an ideal cycle, which is 4K wide and centred round $T_m$. We observe that the maximum magnetic work is attained around the critical temperature, and it tends to zero at higher temperature due to the low contribution of the paramagnetic phase. Expanding the temperature span enhances the output magnetic work, and the same effect can be achieved by increasing the magnetic field span (Supplementary Material).

These simulations highlight the need for materials that exhibit a magnetic transition exactly within the operating temperature range of the TM device. This range is defined not only by the temperatures of the thermal sources but also by the effective heat transferred during the cycle.

### 3.2 In-operando test of materials

The TM performance of the three Heusler alloys was directly tested in the small-scale TM motor prototype. A mass of 0.23±0.01 g of the three materials was shaped into a ring rotor and mounted in the TM motor. The torque, angular speed and mechanical power output of the rotor, generated by exploiting TM cycles, were measured by varying the temperature of warm source ($T_1$) and the external resistive load.

Figure 7 reports the mechanical power output as a function of the angular speed for the three materials at different values of $T_1$, while maintaining the cold reservoir temperature constant at 297 K. Generally, the power output increases as the temperature of the warm heat source increases, owing to the enlargement of the cycle area. However, this trend varies with the rotation speed. At a high external load, the rotation speed is low, limiting the power output. As the load decreases, the rotational speed increases, leading to an increase of the power output to a maximum value. With a further increase in speed, we observe a stabilization or even a decrease in power output, especially for measurements performed with $T_1$ exceeding the Curie temperature of the exploited material. This effect is caused by the reduction in the heat exchanged during the cycle. This results in a decrease in the cycle area and, consequently, in the output power. When the mean temperature of the cycle surpasses the Curie temperature of the active TM material, the reduction in *dM/dT* induces a decrease in the net torque of the rotor and, consequently, in the power output. This effect is more pronounced for NiMnCuGa (Figure 7.c), which has the lowest Curie temperature in the series.

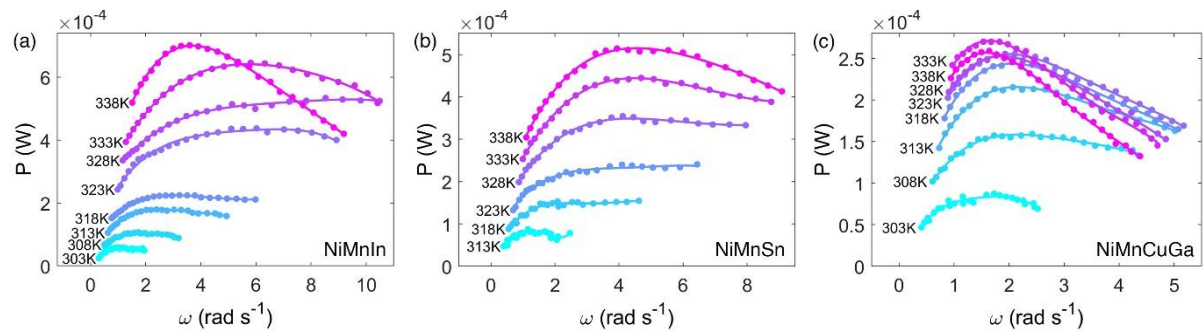

*Figure 7: Mechanical power output of the TM motor prototype as a function of the shaft speed obtained by varying the temperature of the warm source ($T_1$) and the external resistive load exploiting rings of (a) NiMnIn, (b) NiMnSn and (c) NiMnCuGa. Each point represents the average between the measurements performed with increasing and decreasing load sweep. The solid lines are the polynomial interpolation of experimental data.*

Figure 8.a compares the maximum power output as a function of $T_1$ for the three materials. NiMnCuGa induces the maximum output at low temperatures. As the warm reservoir temperature increases, the NiMnCuGa output tends to stabilize, whereas the output of NiMnIn and NiMnSn increases. The NiMnIn alloy shows a maximum mechanical output of 0.70±0.04 mW, which corresponds to 3.0±0.1 W/kg or 23.7±0.4 mW/cm$^3$, considering the output power normalized to the mass or to the volume of the active material. Moving from the In-based to the Sn-based alloy, there is a reduction of about 26% in maximum output power, consistent with the results from the calculations of the cycles based on $M(H)$ data.

The temperature-dependent power output assessed through the evaluation of materials in the TM motor, mirrors the temperature-behaviour of the magnetic work of the thermomagnetic cycle shown in Figure 6 calculated from $M(H,T)$ data considering the temperatures of the cold and warm reservoirs as the boundaries of the cycle. Nevertheless, a comparison between the experimental power and estimated magnetic work of the cycle reveals a shift of the experimental curves toward higher temperatures. The maximum power output for NiMnIn and NiMnSn continues to increase with temperature through the 300-340 K range, whereas, the calculated magnetic work of the cycle almost saturates between 320 and 330 K. Similarly, the output measured using NiMnCuGa reaches saturation approximately 10 K above the calculated curve. This result confirms the hypothesis that the cycle effectively followed by the material does not reach the temperature of the heat sources. This effect, attributed to the slow heat exchange of the thermomagnetic material with the heat sources, limits the power obtainable by the TM motor. Indeed, it is evident that the directly measured power output values are lower than the magnetic work calculated from thermomagnetic cycles (Figure 6.a) when multiplied by the frequency of rotation. For instance, in the case of NiMnIn, the maximum power output is 3.0±0.1 W/kg at a frequency of 0.6 Hz, equivalent to a mass-normalized work per cycle of 5.0±0.3 J/kg. This value significantly contrasts with the 21 J/kg calculated for a thermomagnetic cycle between 297 K and 338 K. Instead, it aligns closely with the maximum calculated value (4.8 J/kg) for a cycle centred at 314 K and characterized by an overall temperature difference of only 4 K (Figure 6.b). This comparison suggests that the limited heat transfer between the active material and the thermal sources prevents the material to reach the temperatures $T_0$ and $T_1$ during its rotation, constraining the TM cycle and reducing the achievable output work. It is important to note that the reported estimations already account for the correction due to friction energy losses (SM) and that magnetic losses can be neglected due to the very low coercivity of the studied materials (see Figure S2 of SM and Supplementary Information of Ref. [29])

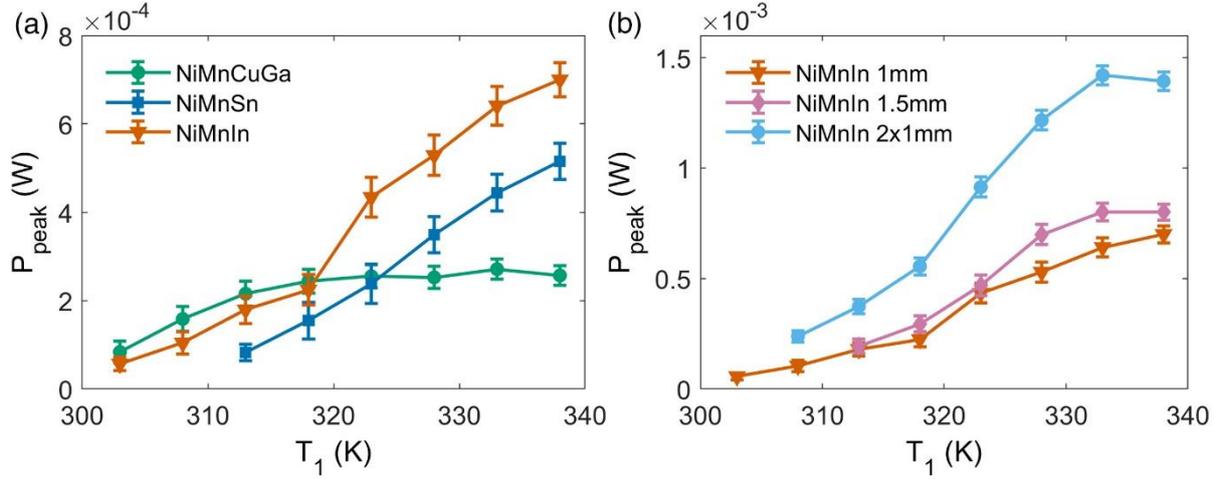

*Figure 8: (a) Maximum mechanical power output as a function of temperature of the warm reservoir for a ring of NiMnIn (orange triangles), NiMnSn (blue squares) and NiMnCuGa (green circles). (b) Comparison of the maximum mechanical power output as a function of temperature of the warm reservoir for: a single ring of NiMnIn with a radial thickness of 1 mm (orange triangles), a single ring of NiMnIn with a radial thickness of 1.5 mm (pink diamonds) and two rings of NiMnIn with a radial thickness of 1 mm fixed to the same shaft (azure circles). Solid lines serve as guide to the eyes.*

To delve this point, we compared, in Figure 8.b, the output of 3 rotors made by the same NiMnIn material: a ring rotor with a radial thickness of 1 mm (internal diameter of 18 mm), a ring rotor with a radial thickness of 1.5 mm (internal diameter of 17 mm) and a tandem-rotor made of 2 rings with a radial thickness of 1 mm on the same shaft. The distance between the two rings in the tandem-rotor, set at 6 mm, is sufficient to treat the two rotors as independent thermal bodies and avoid any effects related to water capillarity. The three rotors contain 0.23±0.01 g, 0.34±0.01 g and 0.46±0.02 g of the active TM material, respectively. The power output of the tandem rotor is approximately double of the single 1 mm rotor, resulting in a mass-normalized power output of 3.0±0.1 W/kg at 338 K. On the contrary, the thickest rotor does not show an increase of the power output proportional to the mass of active magnetic material: its normalized power is limited to 2.4±0.1 W/kg. This limitation can be attributed to the heat transfer between the rotor and the environment. Indeed, we observe that the mass-normalized power output of the three rotors (Figure S4 of SM) scales well with their surface-to-volume ratio, which is 4 mm$^{-1}$ and 3 mm$^{-1}$ for the 1 mm rotors and 1.5 rotor, respectively.

This result confirms that the heat exchange at the interface between the rotor and the fluid (air or water) is currently the limiting factor of the power output of the TM motor. Enhancing heat exchange at the interface and increasing the surface-to-volume ratio of the active rotor are critical objectives for further improving the power and efficiency of TMGs.

To further investigate this aspect, it is imperative to determine the effective temperatures reached by the active material during the cycle and how they vary throughout rotation. Future work will involve conducting new experiments, thermographic analysis, and finite element simulations of the complex thermodynamic system to achieve this objective. These results will also facilitate the reliable estimation of heat absorbed by the active material from the warm thermal source, thereby enabling the calculation of efficiencies in the thermomechanical energy conversion process.

A preliminary result of infrared (IR) thermography analysis of the rotor during operation is shown in Figure 9. The thermal image was acquired using a high-resolution infrared camera (FLIR SC7000) while the TM motor, equipped with the 1 mm thick NiMnIn rotor, operated with the warm source temperature set to 323 K and a rotational speed of 1.5 rad/s. The thermal image of the TM motor (Figure 9.a) and the analysis of the angular temperature distribution in the rotor (Figure 9.b) reveal a thermal gradient in the portion of the rotor outside the water. A total temperature difference of about 6.5 K is observed, with the maximum temperature being 5 K lower than the water temperature and the minimum temperature of the cycle being about 14 K higher than room temperature. This preliminary thermal analysis supports the hypothesis derived from comparing experimental data and magnetic work calculations: 1) the temperature difference experienced by the material within the cycle is lower than the difference between the thermal sources, and 2) the maximum temperature reached by the material is lower than the temperature of the warm source.

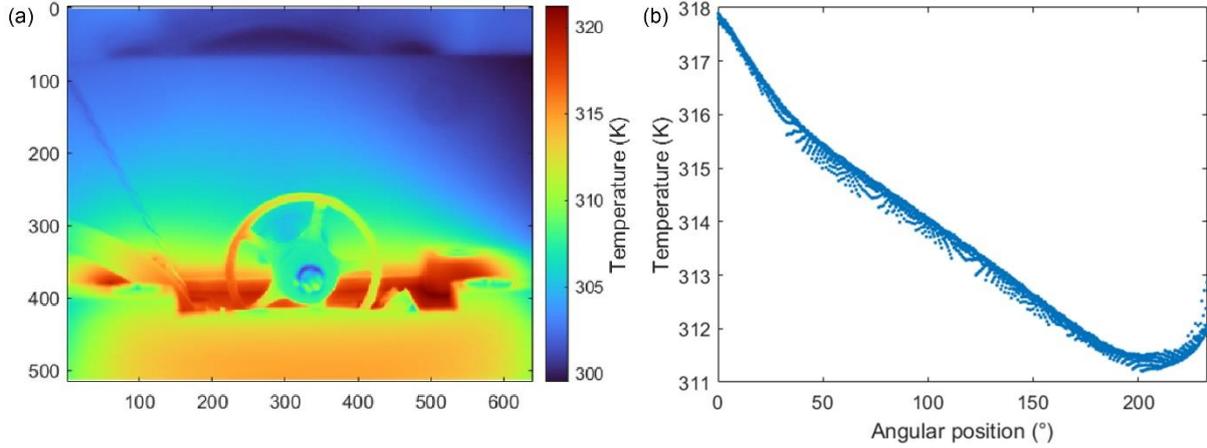

Figure 9: (a) Infrared thermal image of the TM motor during operation. The motor is equipped with the 1 mm thick NiMnIn rotor, and the warm heat source is maintained at 323 K. The load is adjusted to achieve a rotational speed of approximately 1.5 rad/s. (b) Temperature of the angular segments of the rotor above the warm water as a function of angular position starting from the water surface on the left.

Regrading the performance of the TM motor prototype, to benchmark our results against the state of the art in TM generators, we estimated the electric power output achievable by connecting the thermomagnetic motor to an electric generator with known characteristics. We considered a miniature two-phase electric generator, optimized for low-rpm operation, featuring a speed constant ($K_v$) of 2 rads$^{-1}$V$^{-1}$ and a winding internal resistance ($R_i$) of 1300 Ω. The potential net electrical power delivered to a load ($P_{load}$) has been calculated using the following equation starting from the measured mechanical power output ($P_{in}$):

$$P_{load} = P_{in}\left(1 - \frac{P_{in}}{\omega^2} \cdot K_v^2 \cdot R_i\right) \quad (6)$$

Detailed information on the estimation of electric power is provided in Paragraph 4 of the SM. Using the tandem-rotor of NiMnIn, the TM motor is expected to yield a maximum electric power output of 1.2 mW at an angular speed of 7.3 rads$^{-1}$. Normalizing this value to the mass or the volume of the utilized active magnetic material, we derived an output electric power of 2.6 W/kg or 20.4 mW/cm$^3$, respectively. Notably, this estimated electric power significantly surpasses that achieved by a similar Curie wheel using a Gd ring (0.45 mW/cm$^3$), as reported by R. A. Kishore et al. [15]. Similarly, static TMG prototypes utilizing Gd and LaFeSi-based materials have exhibited average power densities below 0.3 mW/cm$^3$. [39,40] This finding underscores the potential of our TM motor prototype to serve also as an efficiently test-bench for optimizing TMGs design, aiming to develop efficient and powerful devices.

## 4. Conclusions and Perspective

This study investigated the potential of three representative Ni,Mn-based Heusler alloys for applications in thermomagnetic harvesting of low-grade waste heat in the temperature range of 300 – 340 K. Their thermomagnetic performance was evaluated by determining the magnetic work of ideal thermomagnetic cycles, as defined by magnetization data as a function of temperature and applied magnetic field. Additionally, their performance was directly measured in a small-scale prototype of thermomagnetic motor, specially designed for testing small masses of TM materials. The mechanical power output of the motor, operated with $Ni_{48}Mn_{36}In_{16}$, $Ni_{48}Mn_{36}Sn_{16}$ and $Ni_{50}Mn_{19}Ga_6Cu_{25}$ alloys was measured by varying the temperature of the warm heat source and the external mechanical load. The $Ni_{48}Mn_{36}In_{16}$ alloy exhibited the highest power output (0.70±0.04 mW), attributed to its high saturation magnetization and its critical temperature falling within the midpoint of the investigated temperature range. $Ni_{48}Mn_{36}Sn_{16}$ displayed a temperature dependence of the output power like the In-alloy but reduced by the 25% and shifted to higher temperature due to its lower saturation magnetization and higher Curie temperature. The $Ni_{50}Mn_{19}Cu_6Ga_{25}$ alloy, while having a lower maximum power compared to the other two alloys, demonstrated larger values of output below 315 K, thanks to its lower Curie temperature.

These results provide an experimental evidence that the performance of a TM energy conversion process is strongly influenced by the magnetic properties of the active magnetic material, specifically in the case of second-order magnetic transitions its Curie temperature and saturation magnetization. This finding highlights the significance of tuning and optimizing the magnetic properties of materials within specific temperature ranges for enhancing their performance in TM applications. In this context, Heusler alloys have proven to be highly promising materials, benefiting from their exceptional tunability of their magnetic properties.

From an experimental standpoint, this work introduces a novel thermomagnetic tester designed for in-operando evaluation of TM materials performance under various conditions. This small-scale prototype of thermomagnetic motor allows for testing small sample masses (less than 0.3 g), typical of laboratory-scale experiments, and serves as a crucial bridge between fundamental material research and prototypes development. The tester can measure the mechanical output energy of the TM motor as a function of the thermal sources temperatures and of the rotational speed, which is varied by applying an external mechanical load. The variation of thermal sources temperature and rotation speed, which control heat exchanges, allows to experimentally realize TM cycles with different widths and evaluate the optimal thermomagnetic performance of each material. This is important to test materials characterized by different magnetization gradients with temperature and materials undergoing a first-order magnetic transition characterized by hysteresis and irreversibility effects. In general, all promising thermomagnetic materials, that can be reduced to a coarse powder (size < 200 µm ) can be implemented and test in the prototype. This is made possible by a straightforward method for producing small ring-rotors, which relies on epoxy-based composites and 3D-printed soluble moulds. The capability of the TM tester to continuously monitor the rotational speed and torque of the rotor enables the indirect assessment of the homogeneity of the prepared rotors and the verification of potential long-term deterioration effects. The remarkable agreement between experimental results, obtained by testing the three materials, and the output magnetic work of ideal thermomagnetic cycles calculated from magnetization data, validated the capability of the developed tester to directly assess thermomagnetic performance of materials decoupled from the efficiency of electrical energy conversion.

Moreover, the estimation of the electric power output achievable by connecting the TM motor to an electric generator underscored the potential of the realized TM energy converter design. The estimated electric power output significantly exceeds that achieved in other Curie wheels and static TMGs. To further enhance the performance of the TM motor it is necessary to independently test and optimize its various components (geometry and material of the active rotor, magnetic field source, heat exchange medium and spatial distribution of the thermal gradient, mechanical-to-electric energy converter). This objective can be accomplished thanks to the flexibility and modularity of the presented prototype, which can also serve as a test-bench for TMG designs. In this work, we present a preliminary study on the geometries of the active elements. We examined three configurations of the active element, revealing that, with characteristics length scales of 1 mm, the bottleneck restricting the output mechanical power of the thermomagnetic motor is the heat exchange at the interface with the transfer fluid. This result underscores the need for further investigations to design active elements with an increased surface to volume ratio, thereby enhancing the heat exchange.

A critical aspect to address in future work is the accurate determination of the temperatures reached by the material during the cycle. This measurement is important for completing the in-operando characterization of the active material, allowing to obtain the temperature boundaries of the TM cycle effectively carried out during operation. Achieving this will optimize the geometry and thermal properties of the thermomagnetic generator, enhancing the thermomagnetic performance of the material. Additionally, it will enable the accurate calculation of the heat absorbed by the magnetic material within the cycle and the efficiency of the thermomagnetic energy conversion process. Future experiments, including thermography analysis and finite-element simulations of the thermodynamic system, will be conducted to achieve these objective.

Moreover, each element of the prototype (i.e.: the magnets assembly, the relative position of water level and field gradient, the electric generator) have to be independently studied, optimized and tested in order to improve the efficiency of the device and increase the fraction of magnetic energy that is converted in mechanical, and electrical, energy.

## Supplementary Material

The Supplementary Material (SM) provides supplementary data from experiments and calculations of magnetic work, along with detailed insights into the methodology used to calculate the mechanical power output of the TM motor and estimate the electrical power output.

## Author statement

**F. Cugini:** Conceptualization, Methodology, Investigation, Formal Analysis, Writing – Original Draft, Writing – Review & Editing, Funding acquisition; **L. Gallo:** Investigation, Writing - Original Draft, Visualization; **G. Garulli**: Investigation, Formal Analysis, Visualization, Software, Writing - Original Draft; **D. Olivieri**: Investigation, Formal Analysis, Software; **G. Trevisi**: Investigation, Resources; **S. Fabbrici**: Resources, Writing - Review & Editing; **F. Albertini**: Supervision, Resources; **M. Solzi**: Supervision, Resources, Writing - Review & Editing, Funding acquisition.

## Declaration of Competing Interest

The authors declare that there is no conflict of interest.

## Acknowledgment

*The authors acknowledge F. Bozzoli, L. Cattani, and M. Malavasi from the Engineering for Industrial Systems and Technologies Department of Parma University for their invaluable assistance with IR thermography. This work was mainly founded by University of Parma through the action Bando di Ateneo 2021 per la ricerca co-funded by MUR-Italian Ministry of Universities and Research - D.M. 737/2021 - PNR - PNRR – NextGenerationEU. This Project was partially funded by European Union – NextGenerationEU under the National Recovery and Resilience Plan (NRRP), Mission 4 Component 2 Investment 1.1 - Call for tender No. 1409 of 14-09-2022 of Italian Ministry of University and Research (Project Code P2022KMXBL, Concession Decree No. 0001381 of 01/09/2023 adopted by the Italian Ministry of Universities and Research, CUP D53D23019360001, "Small-scale Thermomagnetic Energy harvesters: from materials to devices"). G. Garulli doctorate scholarship is founded by European Union – NextGenerationEU, through the DM 351/2022 - PNRR – NextGenerationEU (Mission 4 Component 1 – "Research PNRR"). This work was partially financed by the European Union -NextGenerationEU (National Sustainable Mobility Center CN00000023, Italian Ministry of University and Research Decree no. 1033, 17/06/2022, Spoke 11, Innovative Materials & Lightweighting). The opinions expressed are those of the authors only and should not be considered as representative of the European Union or the European Commission's official position. Neither the European Union nor the European Commission can be held responsible for them.*

# In-operando test of tunable Heusler alloys for thermomagnetic harvesting of low-grade waste heat

F. Cugini[1,2]*, L. Gallo[1,2], G. Garulli[2], D. Olivieri[1], G. Trevisi[2], S. Fabbrici[2], F. Albertini[2], M. Solzi[1,2]

[1]Department of Mathematical, Physical and Computer Sciences, University of Parma, Parco Area delle Scienze 7/A, 43124 Parma, Italy
[2]Institute of Materials for Electronics and Magnetism - National Research Council (IMEM-CNR), Parco Area delle Scienze 37/A, 43124 Parma, Italy
*francesco.cugini@unipr.it

# Supplementary Material

## 1. Permanent magnet assembly of the thermomagnetic motor

The assembly of permanent magnets used in the thermomagnetic motor consists of three NdFeB magnets, each measuring 5x5x20 mm. The magnets are arranged in the configuration schematized in Figure S1.b. The stray magnetic field generated by the assembly was measured using a SENIS® 3DHALL sensor mounted on a 3D moving system. Figure S1.a presents a map of the stray magnetic field at 1 mm from the surface of the magnet assembly.

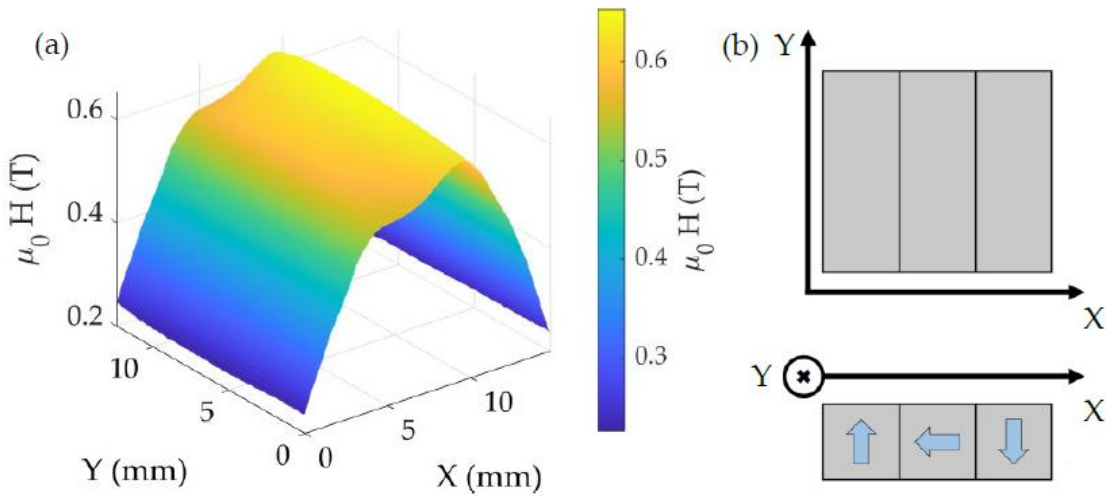

*Figure S1 (a) Strength of the stray magnetic field generated at 1 mm from the surface of permanent magnets assembly schematized in figure (b). The light blue arrows, in figure (b), represent the direction of the magnetic axes of each permanent magnets.*

## 2. **Hysteresis loops of NiMnIn, NiMnSn and NiMnCuGa samples**

Hysteresis loops were collected at room temperature (291-293 K) for the three Heusler samples with an extraction magnetometer.

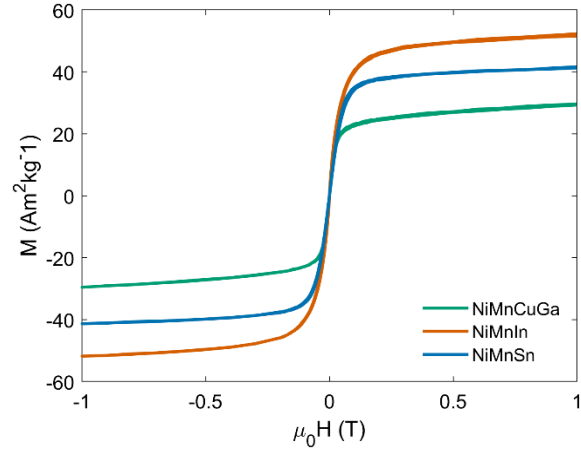

*Figure S2: hysteresis loops collected at room temperature for $Ni_{48}Mn_{36}Sn_{16}$ (blue line), $Ni_{48}Mn_{36}In_{16}$ (orange line), and $Ni_{50}Mn_{19}Cu_6Ga_{25}$ (green line) samples.*

## 3. **Estimation of magnetic work from TM cycles**

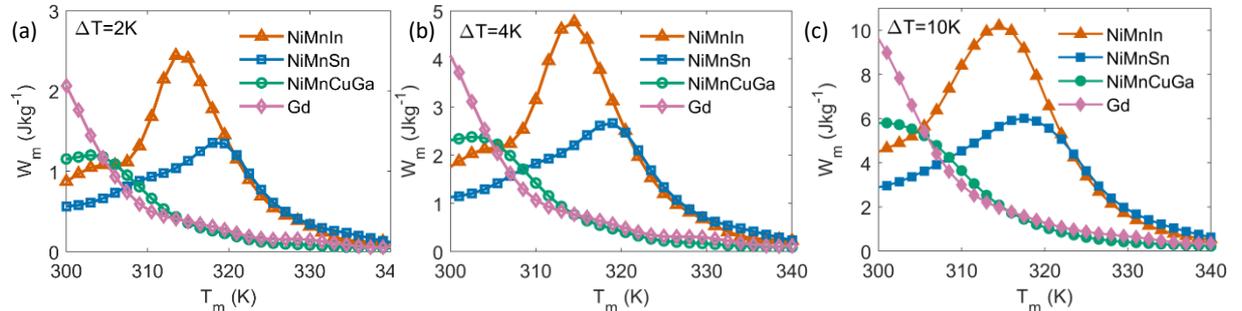

*Figure S3: Useful magnetic work of an ideal TM cycle operated with NiMnIn, NiMnSn, NiMnCuGa and Gd materials as a function of the mean temperature of the cycle Tm, with a temperature span between the cold and warm source fixed at (a) 2 K, (b) 4K and (c) 10 K. The magnetic work is calculated through eq. 1 from M(T,H) data reported in Figure 5 of the main text.*

## 4. Direct test of materials in the thermomagnetic motor prototype

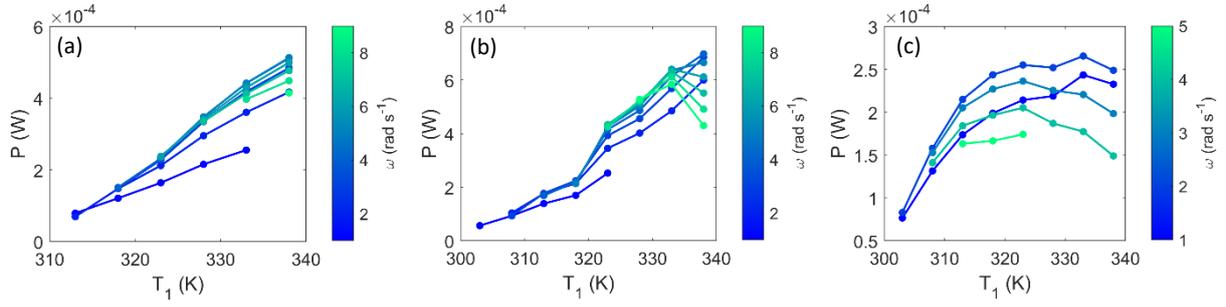

*Figure S4: mechanical power output of the TMG prototype as a function of the warm source temperature ($T_1$) at fixed shaft speeds for rings of (a) NiMnIn, (b) NiMnSn and (c) NiMnCuGa of active TM material. Each point was obtained by interpolation of the power vs speed measurements at fixed warm source temperature ($T_1$).*

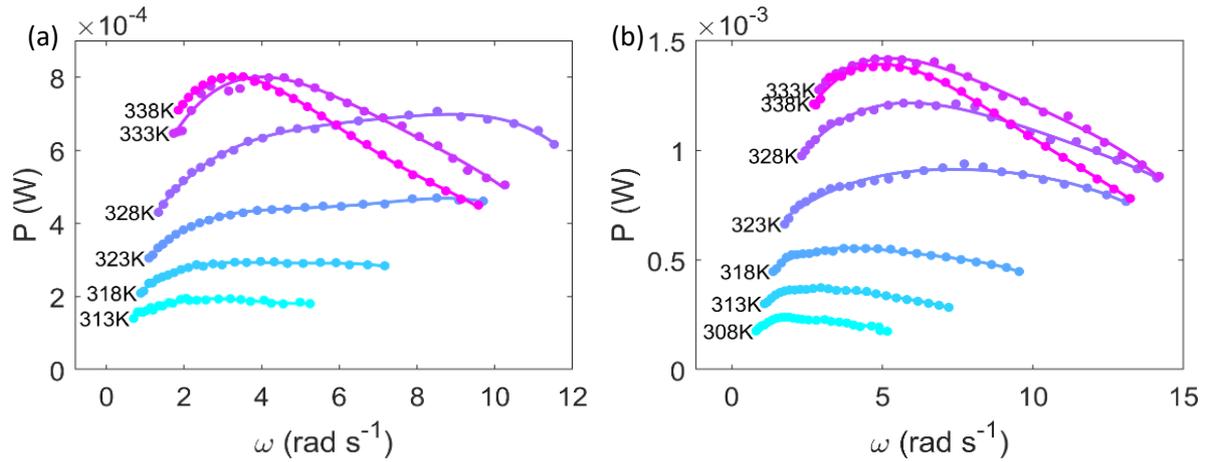

*Figure S5: mechanical power output of the TMG prototype as a function of the shaft speed obtained by varying the temperature of the warm source ($T_1$) and the external resistive load exploiting (a) a ring of NiMnIn with a thickness of 1.5 mm and (b) two rings of NiMnIn with a thickness of 1.0 mm connected to the same shaft. Each point represents the average between the measurements performed with increasing and decreasing load sweep. The solidi lines are the polynomial interpolation of experimental data.*

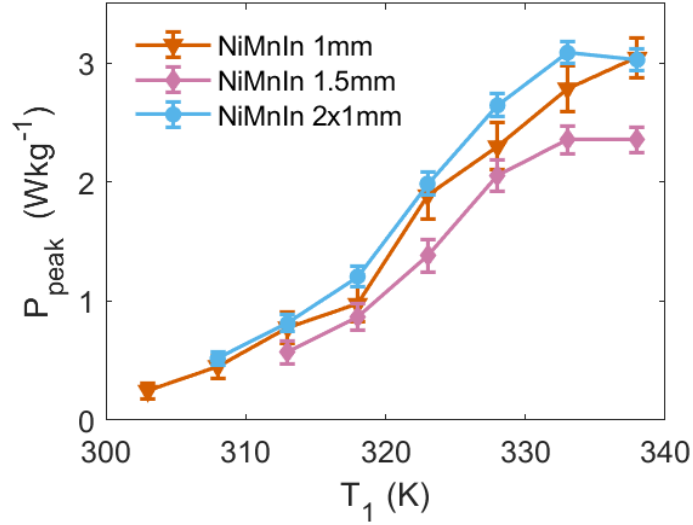

*Figure S6: Comparison of maximum mechanical power output, normalized to the mass of active magnetic material, as a function of temperature of the warm reservoir for: a single ring of NiMnIn with a thickness of 1 mm (triangles), a single ring of NiMnIn with a thickness of 1.5 mm (diamonds) and two rings of NiMnIn with a thickness of 1 mm fixed to the same shaft (circles). Solid lines serve as guide to the eyes.*

## 5. Data acquisition and calibration of the thermomagnetic motor

The mechanical power, angular speed and torque of the thermomagnetic motor were obtained by measuring voltage and current output of a custom-designed two-phase electric generator connected to the rotor's shaft. Voltage was directly measured at the resistive load, while current was recorded by detecting the potential difference across a 100 Ω shunt resistor.

In a permanent magnet electric motor or generator, the open circuit back-emf, which in our case has near-sinusoidal shape, exhibits a peak voltage value ($V_{peak}$) that is directly proportional to the angular speed ($\omega$). The proportionality constant, known as the motor speed constant ($K_v$), is defined as:

$$K_v = \frac{\omega}{V_{peak}} \quad (1)$$

Eq. (1) can be used to estimate the angular speed from the peak voltage value, knowing the motor speed constant. However, to obtain $\omega$ at each recorded time instant, two windings (called A and B), phased out by 90 electrical degrees, are needed. Each winding can be modelled as an ideal sinusoidal voltage generator, producing a time-dependent signal $V_{idealA,B}(t)$, in series with a resistor of value equal to the internal resistance of the winding ($R_{iA}$, $R_{iB}$). If we applied a load to the A and B circuits, the back-emf of each winding as a function of time can be expressed as:

$$V_{idealA}(t) = V_{loadA}(t) + R_{iA} \cdot I_{loadA}(t) \quad (2)$$

$$V_{idealB}(t) = V_{loadB}(t) + R_{iB} \cdot I_{loadB}(t) \quad (3)$$

The back-emf can be calculated from the voltage at the load corrected for resistive losses in the windings ($R_{iA,B} \cdot I_{loadA,B}(t)$). Considering Eq. (1) and assuming a sinusoidal voltage output with an angular frequency $\omega$ and a pure resistive load, the back-emf of each winding as function of time can also be modelled as:

$$V_{idealA}(t) = V_{peakA} \sin(\omega t + \varphi) = \frac{\omega}{K_{vA}} \sin(\omega t + \varphi) \qquad (4)$$

$$V_{idealB}(t) = V_{peakB} \cos(\omega t + \varphi) = \frac{\omega}{K_{vB}} \cos(\omega t + \varphi) \qquad (5)$$

It is possible to show that by computing the quadrature sum of $K_{vA}V_{idealA}(t)$ and $K_{vB}V_{idealB}(t)$ we can estimate the angular speed $\widetilde{\omega}(t)$ d for any time instant t:

$$\omega(t) = \sqrt{K_{vA}^2 \cdot V_{idealA}^2(t) + K_{vB}^2 \cdot V_{idealB}^2(t)} =$$

$$\sqrt{K_{vA}^2 \left(\frac{\omega}{K_{vA}} \sin(\omega t + \varphi)\right)^2 + K_{vB}^2 \left(\frac{\omega}{K_{vB}} \sin(\omega t + \varphi)\right)^2} = \sqrt{\omega^2 \sin^2(\omega t + \varphi) + \omega^2 \cos^2(\omega t + \varphi)} =$$

$$\omega\sqrt{\sin^2(\omega t + \varphi) + \cos^2(\omega t + \varphi)} = \omega \qquad (6)$$

By substituting Eq.s 2 and 3 in Eq. 6, we can determine the angular speed of the shaft using the output voltage ($V_{load}(t)$) and current ($I_{load}(t)$) also when load is applied to the generator:

$$\omega(t) = \sqrt{K_{vA}^2 \cdot \left(V_{loadA}(t) + R_{iA} \cdot I_{loadA}(t)\right)^2 + K_{vB}^2 \cdot \left(V_{loadB}(t) + R_{iB} \cdot I_{loadB}(t)\right)^2} \qquad (7)$$

Different motor speed constants ($K_{vA}$ and $K_{vB}$) and internal resistance values ($R_{iA}$ and $R_{iB}$) are used for each winding to account for minor discrepancies arising from tolerances during the manufacturing process of the generator stator. The series resistance of each winding was directly measured using a digital multimeter. Whereas a calibration process was required to precisely determine the speed constants. During calibration, voltage and current values were recorded while applying a known torque to the generator shaft with various applied electric loads. A one-term Fourier series fit was utilized to determine both the rotational speed and the amplitude of the voltage output after corrections for internal resistance. With the obtained values of $\omega$ and voltage output, a linear fit was performed to obtain the value of $K_v$ for each winding.

Similarly to Eq. 1, the relation between peak armature current ($I_{peak}$) and the torque ($\tau$) exerted by the motor/generator is:

$$K_\tau = \frac{\tau}{I_{peak}} \qquad (8)$$

where $K_\tau$ is the motor torque constant. To eliminate detent torque in the generator, a 3D printed ironless stator core was used to support the windings. For a permanent magnet-based motor or generator with ironless core, when the motor speed constant and torque constant are expressed in SI units, the motor speed constant equals the inverse of the torque constant ($K_\tau = \frac{1}{K_v}$) [Hamid A. Toliyat, G. B. Kliman, Electrical and computer engineering, 120, *Handbook of electric motors, 2nd ed., rev. and exp, chapter 2.3.9*]. Therefore, the torque of the motor can be calculated as:

$$\tau = \frac{I}{K_v} \qquad (9)$$

This equation allows us to estimate the motor torque with the same approach used in eq. 7 by measuring the winding current and knowing the speed constant $K_v$ of each winding:

$$\tau(t) = \sqrt{\frac{I^2_{loadA}(t)}{K_{vA}^2} + \frac{I^2_{loadB}(t)}{K_{vB}^2}} \qquad (10)$$

During the calibration process, the difference between the applied known torque and the torque inferred from armature current was calculated to measure the friction torque caused by the ball bearings supporting the shaft. The torque caused by friction was estimated to be approximately $2 \cdot 10^{-5}$ Nm and showed no strong dependence on $\omega$. If 1 mW of mechanical power is applied to the shaft at a speed of 5 rads$^{-1}$, the friction losses would correspond to approximately 10% of the mechanical power input. The torque data acquired during the test of materials are corrected for the measured values of friction torque to obtain the overall torque generated by the thermomagnetic motor.

## 6. Electric power generation and efficiency calculations

With the knowledge of the mechanical power output versus angular speed curve of the thermomagnetic motor, it is possible to simulate the electric power output and efficiency of conversion achievable by connecting the motor to an electric generator with a known speed constant and internal resistance. In a two-phase electric generator, assuming negligible friction, constant speed and that the two windings have the same speed constant $K_v$ and internal resistance $R_i$, the total power loss due to joule effect in the generator armatures is given by:

$$P_{loss} = I^2_{peak} \cdot R_i = \tau^2 \cdot K_v^2 \cdot R_i = \frac{P_{in}^2}{\omega^2} \cdot K_v^2 \cdot R_i \qquad (11)$$

where $P_{in}$ is the mechanical power input and $\omega$ is the angular speed.

The overall conversion efficiency ($\eta$) from mechanical to electrical energy can be defined as follows:

$$\eta = 1 - \frac{P_{loss}}{P_{in}} \qquad (12)$$

The generator efficiency ($\eta$) and the net electrical power delivered to the load ($P_{load}$) can thus be estimated with respect to mechanical power input, angular speed and generator construction parameters:

$$\eta = 1 - \frac{P_{in}}{\omega^2} \cdot K_v^2 \cdot R_i \qquad (13)$$

$$P_{load} = P_{in} \cdot \eta = P_{in} \left(1 - \frac{P_{in}}{\omega^2} \cdot K_v^2 \cdot R_i\right) \qquad (14)$$

Assuming plausible values of the speed constant $K_v$ = 2 rads$^{-1}$V$^{-1}$ and winding internal resistance $R_i$ = 1300 Ω for a miniature electric generator optimized for low-rpm operation, we can estimate the electric power output of a thermomagnetic generator using the materials tested in this work. The two-rotor NiMnIn-based configuration of the thermomagnetic motor is expected to give a maximum electric power output of 1.2 mW at an angular speed of 7.3 rads$^{-1}$, corresponding to a mechanical-to-electric power conversion efficiency of 86%. If we normalize this value to the mass ore the volume of utilized active magnetic material, we obtained an output electric power of 2.6 W/kg or 20.4 mW/cm$^3$.